\documentclass[aps,showpacs,floatfix,nofootinbib]{revtex4-2}
\usepackage{amsmath,amssymb,graphicx,bm}

\begin{document}

\title{Nambu dynamics and its noncanonical Hamiltonian representation \\
in many degrees of freedom systems}


\author{Atsushi Horikoshi}\email{horikosi@tcu.ac.jp}
\affiliation{Department of Natural Sciences, Tokyo City University,Tokyo 158-8557, Japan}


\begin{abstract}%
Nambu dynamics is a generalized Hamiltonian dynamics of more than two variables, 
whose time evolutions are given by the Nambu bracket, a generalization of the canonical Poisson bracket.
Nambu dynamics can always be represented in the form of noncanonical Hamiltonian dynamics
by defining the noncanonical Poisson bracket by means of the Nambu bracket.
For the time evolution to be consistent,
the Nambu bracket must satisfy the fundamental identity, 
while the noncanonical Poisson bracket must satisfy the Jacobi identity.
However, in many degrees of freedom systems, 
it is well known that the fundamental identity does not hold. 
In the present paper we show that, even if the fundamental identity is violated,
the Jacobi identity for the corresponding noncanonical Hamiltonian dynamics could hold.
As an example we evaluate these identities for a semiclassical system of two coupled oscillators.
\end{abstract}


\maketitle
\section{Introduction}
\label{Introduction}
There are various ways to generalize the Hamiltonian dynamics.
In the present paper,
we focus on two generalized dynamics, the Nambu dynamics and the noncanonical Hamiltonian dynamics.
The Nambu dynamics is a generalized Hamiltonian dynamics that is defined in the extended phase space 
spanned by $N(\ge 3)$ variables $(x_1, x_2, ..., x_{N})$ \cite{Nambu}.
Taking the Liouville theorem as a guiding principle,
Nambu generalized the Hamilton equations of motion to the Nambu equations, 
which are defined by $N-1$ Hamiltonians and the Nambu bracket, 
an $N$-ary generalization of the canonical Poisson bracket.
In order for the variable transformation including the time evolution to be consistent, 
the Nambu bracket must satisfy the fundamental identity,
a generalization of the Jacobi identity \cite{SahooValsakumar1,SahooValsakumar2,Takhtajan}. 
On the other hand, the noncanonical Hamiltonian dynamics is also defined in the $N$-dimensional extended phase space,
and the Hamilton equations of motion are generalized to the noncanonical ones,
which are defined by one Hamiltonian and the noncanonical Poisson bracket \cite{Morrison}.
Although the noncanonical Poisson bracket has the same structure as the canonical Poisson bracket, 
it is defined by means of the variable-dependent $N \times N$ Poisson matrix.
The noncanonical Poisson bracket must satisfy the Jacobi identity 
for the consistent variable transformation including the time evolution. 
It has been shown that the Nambu dynamics can always be represented 
in the form of the noncanonical Hamiltonian dynamics with the noncanonical Poisson bracket 
defined by the Nambu bracket \cite{Takhtajan,BialynickiBirulaMorrison}.
\par
Although the structure of the Nambu dynamics has impressed many authors,
it has been revealed that the Nambu bracket exhibits serious difficulties 
in many degrees of freedom systems \cite{Nambu,SahooValsakumar1,SahooValsakumar2,Takhtajan,HoMatsuo}.
This is because in such systems 
the Nambu bracket does not satisfy the fundamental identity.
Since the fundamental identity is too strict,
each degree of freedom must be decoupled to satisfy the identity.
On the other hand, for the noncanonical Poisson bracket,
whether or not the Jacobi identity holds is not a matter of the number of degrees of freedom, 
but rather a matter of the nature of the Poisson matrix. 
\par
In the present paper, 
we study the Nambu dynamics and the corresponding noncanonical Hamiltonian dynamics
in many degrees of freedom systems,
and show that even if the fundamental identity is violated, the Jacobi identity for corresponding dynamics could hold.
That is, even if the consistent time evolution is broken in the Nambu dynamics, 
it could be restored in the corresponding noncanonical Hamiltonian dynamics.
As an example we evaluate these two identities for a simplified H\'enon--Heiles model \cite{HellerStechelDavis},
a system of two coupled oscillators 
whose semiclassical dynamics has been studied using the hidden Nambu formalism \cite{HorikoshiKawamura,Horikoshi}.
\par
The outline of this paper is as follows.
In Sect. 2 we review the Nambu dynamics and its noncanonical Hamiltonian representation
with proofs of the fundamental identity and the corresponding Jacobi identity.
In Sect. 3 we show the violation of the fundamental identity for the Nambu bracket in many degrees of freedom systems,
and give the condition under which the Jacobi identity 
for the corresponding noncanonical Poisson bracket holds. 
We also present an example of a two degrees of freedom system. 
Our conclusions are given in the last section. 

\section{Nambu dynamics and noncanonical Hamiltonian dynamics}
\label{Nambu dynamics and noncanonical Hamiltonian dynamics}
We begin with a brief review of the Nambu dynamics \cite{Nambu} and
the relationship with the noncanonical Hamiltonian dynamics \cite{Takhtajan,BialynickiBirulaMorrison}
in one degree of freedom systems.
Throughout this paper we treat the case of $N=3$, and therefore
we consider the dynamics of three Nambu variables $(x_1, x_2, x_3)$ in this section.
The generalization for arbitrary $N\ge3$ is straightforward. 
 
\subsection{Nambu dynamics}
\label{Nambu dynamics}
In the Nambu dynamics, the canonical Poisson bracket is generalized to the Nambu bracket 
defined by means of the 3D Jacobian,
\begin{eqnarray}
\{A, B, C\}\equiv
\frac{\partial (A, B, C)}{\partial (x_1, x_2, x_3)}
= \epsilon_{i j k} \frac{\partial A}{\partial x_{i}}\frac{\partial B}{\partial x_{j}}\frac{\partial C}{\partial x_{k}},
\label{NB}
\end{eqnarray}
where $A, B$, and $C$ are any functions of the three variables $(x_1, x_2, x_3)$ and 
$\epsilon_{i j k}$ is the 3D Levi--Civita symbol.
We employ the summation convention over repeated indices throughout this paper.
In terms of the Nambu bracket, 
the Nambu equation for any function ${f}={f}(x_1, x_2, x_3)$ can be written as
\begin{eqnarray}
\frac{df}{dt} = \{f, H, G\}
= \epsilon_{i j k} \frac{\partial f}{\partial x_{i}}\frac{\partial H}{\partial x_{j}}\frac{\partial G}{\partial x_{k}},
\label{N-eq}
\end{eqnarray}
where $H$ and $G$ are Nambu Hamiltonians.
The time evolution according to this equation preserves the 3D phase space volume,
and therefore the Liouville theorem holds in the Nambu dynamics.
\par
The Nambu bracket of Eq. (\ref{NB}) must satisfy 
the following fundamental identity \cite{SahooValsakumar1,SahooValsakumar2,Takhtajan}:
\begin{eqnarray}
\{\{A, B, C\}, D, E\}= 
\{\{A, D, E\}, B, C\}+
\{A,\{B, D, E\}, C\}+
\{A, B, \{C, D, E\}\}.
\label{FI}
\end{eqnarray}
Here $D$ and $E$ are any functions of the three variables, 
and play the roles of the generating functions of a variable transformation.
In particular, if we choose them as the Nambu Hamiltonians, $(D, E)=(H, G)$,
then the identity of Eq. (\ref{FI}) means that 
the distributive property of time derivatives holds:
\begin{eqnarray}
\frac{d}{dt}\{A, B, C\}= 
\{\frac{d}{dt}A, B, C\}+
\{A,\frac{d}{dt}B, C\}+
\{A, B, \frac{d}{dt}C\}.
\label{FItime}
\end{eqnarray}
Therefore, if the fundamental identity is violated, the consistent time evolution is broken,
at least in the sense that the distributive property does not hold.\footnote{
The violation of the Jacobi identity also implies the breaking of the consistent time evolution.
It is an interesting subject to study how the violation of these identities affects the actual dynamics. 
For example, see Ref. \cite{CaliganChandre}.}
\par
The fundamental identity can be proved as follows \cite{SahooValsakumar2}.
The difference between the left-hand side and the right-hand side of Eq. (\ref{FI}) 
can be represented as
\begin{eqnarray}
{\rm lhs-rhs}=-\left(
\epsilon_{i \mu \nu}\epsilon_{\rho j k}
+\epsilon_{i \nu \rho}\epsilon_{\mu j k}
+\epsilon_{i \rho \mu}\epsilon_{\nu j k}
\right)
\partial_{\mu}A~\!\partial_{\nu}B~\!\partial_{\rho} C~\!\partial_i\left(\partial_j D\partial_k E\right),
\label{FI1}
\end{eqnarray}
which can be rewritten in terms of the generalized Kronecker delta,
\begin{eqnarray}
{\rm lhs-rhs}&=&-\frac{1}{2}~\!\delta^{l m n}_{\mu\nu\rho}~\!
\epsilon_{i l m}\epsilon_{n j k}~\!
\partial_{\mu}A~\!\partial_{\nu}B~\!\partial_{\rho} C~\!\partial_i\left(\partial_j D~\!\partial_k E\right)\nonumber\\
&=&-\frac{1}{2}~\!\epsilon_{\mu\nu\rho}\epsilon_{l m n}
\epsilon_{i l m}\epsilon_{n j k}~\!
\partial_{\mu}A~\!\partial_{\nu}B~\!\partial_{\rho} C~\!\partial_i\left(\partial_j D~\!\partial_k E\right)\nonumber\\
&=&-\epsilon_{\mu\nu\rho}\epsilon_{i j k}~\!
\partial_{\mu}A~\!\partial_{\nu}B~\!\partial_{\rho} C~\!\partial_i\left(\partial_j D~\!\partial_k E\right)\nonumber\\
&=&-\epsilon_{\mu\nu\rho}\epsilon_{i j k}~\!
\left(\partial_i\partial_j D~\!\partial_k E+\partial_j D~\!\partial_i\partial_k E\right)\partial_{\mu}A~\!\partial_{\nu}B~\!\partial_{\rho} C
\nonumber\\
&=&0.
\label{FI2}
\end{eqnarray}
Note that we do not distinguish upper and lower indices.
\subsection{Noncanonical Hamiltonian representation}
\label{Noncanonical Hamiltonian representation}
Start with the Nambu equation of Eq. (\ref{N-eq}).
Using one of the Nambu Hamiltonians, $G$, we define the Poisson matrix $J_{ij}(x_1,x_2,x_3)$ as
\begin{eqnarray}
J_{ij} \equiv \epsilon_{i j k} \frac{\partial G}{\partial x_{k}},
\label{PM1}
\end{eqnarray}
which is anti-symmetric: $J_{ji}=-J_{ij}$.
In terms of this matrix, we define the noncanonical  Poisson bracket as
\begin{eqnarray}
\{A, B\}_{G}\equiv J_{i j} \frac{\partial A}{\partial x_{i}}\frac{\partial B}{\partial x_{j}}=\{A, B, G\},
\label{ncPB1}
\end{eqnarray}
where $A$ and $B$ are any functions of $(x_1, x_2, x_3)$.
Then we can rewrite the Nambu equation as the noncanonical Hamilton's equation of motion:
\begin{eqnarray}
\frac{d f}{dt} = \{f, H, G\}=\{f, H\}_{G}
= J_{i j} \frac{\partial f}{\partial x_{i}}\frac{\partial H}{\partial x_{j}}.
\label{NC-eq1}
\end{eqnarray}
The Jacobi identity for the noncanonical  Poisson bracket of Eq. (\ref{ncPB1}) immediately follows 
from the fundamental identity.  
Let $C=G$, $E=G$, and $D=C$ in the fundamental identity of Eq. (\ref{FI}), then we obtain the Jacobi identity:
\begin{eqnarray}
\{\{A, B\}_{G}, C\}_{G}= 
\{\{A, C\}_{G}, B\}_{G}+
\{A,\{B, C\}_{G}\}_{G}.
\label{JI}
\end{eqnarray}
\par
This way of representing the Nambu dynamics in the form of noncanonical Hamiltonian dynamics
is not unique. 
For example, defining another Poisson matrix as
\begin{eqnarray}
\tilde{J}_{ij} \equiv -\epsilon_{i j k} \frac{\partial H}{\partial x_{k}}
\label{PM2}
\end{eqnarray}
and another noncanonical Poisson bracket as
\begin{eqnarray}
\{A, B\}_{H}\equiv \tilde{J}_{i j} \frac{\partial A}{\partial x_{i}}\frac{\partial B}{\partial x_{j}}=\{A, B, H\},
\label{ncPB2}
\end{eqnarray}
we obtain another expression for the equation of motion:
\begin{eqnarray}
\frac{d f}{dt} = \{f, H, G\}=\{f, G\}_{H}
= \tilde{J}_{i j} \frac{\partial f}{\partial x_{i}}\frac{\partial G}{\partial x_{j}}.
\label{NC-eq2}
\end{eqnarray}
The bracket of Eq. (\ref{ncPB2}) also satisfies the Jacobi identity.
Note that the Liouville theorem holds in the dynamics of both Eqs. (\ref{NC-eq1}) and (\ref{NC-eq2}).
\section{Many degrees of freedom systems}
\label{Many degrees of freedom systems}
It is possible to extend the Nambu dynamics to many degrees of freedom systems.
However, in general, the fundamental identity does not hold 
in such systems \cite{Nambu,SahooValsakumar1,SahooValsakumar2,Takhtajan,HoMatsuo}.
Therefore it is nontrivial whether the Jacobi identity for the noncanonical Poisson bracket
defined by the Nambu bracket holds or not. 
Here we give the conditions under which the identities hold.
As an example we evaluate these identities for a semiclassical system of two coupled oscillators.

\subsection{Nambu dynamics}
\label{Nambu dynamics-n}
Consider a system of $3n$ Nambu variables $(x_1^{1}, x_2^{1}, x_3^{1}, ..., x_1^{n}, x_2^{n}, x_3^{n})$. 
Their time evolution can be given in the same form as Eq. (\ref{N-eq})
by extending the definition of the Nambu bracket,
\begin{eqnarray}
\{A, B, C\} \equiv\sum_{{\alpha}=1}^{n}
\frac{\partial (A, B, C)}{\partial (x_1^{\alpha}, x_2^{\alpha}, x_3^{\alpha})}
=\sum_{{\alpha}=1}^{n} \epsilon_{i j k} 
\frac{\partial A}{\partial x^{\alpha}_{i}}\frac{\partial B}{\partial x^{\alpha}_{j}}\frac{\partial C}{\partial x^{\alpha}_{k}},
\label{NBn}
\end{eqnarray}
where $A,B$, and $C$ are any functions of the $3n$ variables.
In terms of this bracket, 
the Nambu equation for any function ${f}={f}(x_1^{1}, x_2^{1}, x_3^{1}, ..., x_1^{n}, x_2^{n}, x_3^{n})$
can be written as
\begin{eqnarray}
\frac{df}{dt} = \{f, H, G\}
= \sum_{{\alpha}=1}^{n} \epsilon_{i j k} 
\frac{\partial f}{\partial x^{\alpha}_{i}}\frac{\partial H}{\partial x^{\alpha}_{j}}\frac{\partial G}{\partial x^{\alpha}_{k}},
\label{N-eqn}
\end{eqnarray}
where $H$ and $G$ are Nambu Hamiltonians.
The Liouville theorem holds as well in this dynamics.
Let us try to prove the fundamental identity for the Nambu bracket of Eq. (\ref{NBn}). 
To simplify the equations, we employ the notation
$\partial A/\partial x^{\alpha}_i=\partial^{\alpha}_i A$.
Using the definition of Eq. (\ref{NBn}), 
the difference between the left- and right-hand sides of Eq. (\ref{FI}) 
can be represented as
\begin{eqnarray}
&&{\rm lhs-rhs}\nonumber\\
&&=-\sum_{{\alpha}=1}^{n}\sum_{{\beta}=1}^{n}\left(
\epsilon_{i \mu \nu}\epsilon_{\rho j k}~\!\partial^{\alpha}_{\mu}A~\!\partial^{\alpha}_{\nu}B~\!\partial^{\beta}_{\rho} C
+\epsilon_{i \nu \rho}\epsilon_{\mu j k}~\!\partial^{\beta}_{\mu}A~\!\partial^{\alpha}_{\nu}B~\!\partial^{\alpha}_{\rho} C
+\epsilon_{i \rho \mu}\epsilon_{\nu j k}~\!\partial^{\alpha}_{\mu}A~\!\partial^{\beta}_{\nu}B~\!\partial^{\alpha}_{\rho} C\right)
\nonumber\\
&&~~~~~~~~~~~~~~~~\times\partial^{\alpha}_i\left(\partial^{\beta}_j D~\!\partial^{\beta}_k E\right).
\label{FI1n}
\end{eqnarray}
Unlike the case of one degree of freedom,
the difference does not vanish in general, but vanishes under some conditions.  
For example, consider the case that $3n$ variables are decoupled in the functions $D$ and $E$,
\begin{eqnarray}
D=\sum_{{\alpha}=1}^{n}D_{\alpha}(x_1^{\alpha}, x_2^{\alpha}, x_3^{\alpha})
,~~~E=\sum_{{\alpha}=1}^{n}E_{\alpha}(x_1^{\alpha}, x_2^{\alpha}, x_3^{\alpha}),
\label{decouple}
\end{eqnarray}
where $D_{\alpha}$ and $E_{\alpha}$ are only functions of  $(x_1^{\alpha}, x_2^{\alpha}, x_3^{\alpha})$.
Then Eq. (\ref{FI1n}) reads
\begin{eqnarray}
{\rm lhs-rhs}=-\sum_{{\alpha}=1}^{n}\left(
\epsilon_{i \mu \nu}\epsilon_{\rho j k}
+\epsilon_{i \nu \rho}\epsilon_{\mu j k}
+\epsilon_{i \rho \mu}\epsilon_{\nu j k}
\right)~\!\partial^{\alpha}_{\mu}A~\!\partial^{\alpha}_{\nu}B~\!\partial^{\alpha}_{\rho} C~\!
\partial^{\alpha}_i\left(\partial^{\alpha}_j D_{\alpha}~\!\partial^{\alpha}_k E_{\alpha}\right).
\label{FI2n}
\end{eqnarray}
We can show that this difference becomes zero in the same way as Eq. (\ref{FI2}).
Although the fundamental identity holds in this case, 
it is almost meaningless as an identity for many degrees of freedom,
because the decomposed $D$ and $E$ as in Eq. (\ref{decouple}) 
mean that there is no interaction between the degrees of freedom.
The functions $D$ and $E$ in Eq. (\ref{FI}) play the roles of the generating functions of
a variable transformation, and in particular they are the Hamiltonians in the time evolution.
Therefore at least one of them must not be decomposed.
\par
If you do not put any conditions on $D$ and $E$, you have to impose restrictions on $A$, $B$, and $C$.
Consider the case that they are functions of a single degree of freedom,
$A=A_{\alpha}$, $B=B_{\alpha}$, and $C=C_{\alpha}$. 
Then the left-hand side of the fundamental identity of Eq. (\ref{FI}) is 
$\{\{A_{\alpha}, B_{\alpha}, C_{\alpha}\}, D, E\}$, 
and Eq. (\ref{FI1n}) reads
\begin{eqnarray}
{\rm lhs-rhs}=-\sum_{{\alpha}=1}^{n}\left(
\epsilon_{i \mu \nu}\epsilon_{\rho j k}
+\epsilon_{i \nu \rho}\epsilon_{\mu j k}
+\epsilon_{i \rho \mu}\epsilon_{\nu j k}
\right)~\!\partial^{\alpha}_{\mu}A_{\alpha}~\!\partial^{\alpha}_{\nu}B_{\alpha}~\!\partial^{\alpha}_{\rho} C_{\alpha}~\!
\partial^{\alpha}_i\left(\partial^{\alpha}_j D~\!\partial^{\alpha}_k E\right).
\label{FI3n}
\end{eqnarray}
The same calculation as in Eq. (\ref{FI2}) shows that this difference becomes zero.
\subsection{Noncanonical Hamiltonian representation}
\label{Noncanonical Hamiltonian representation2}
Similar to the case of one degree of freedom,
the Nambu dynamics of Eq. (\ref{N-eqn}) can be represented in the form of the noncanonical Hamiltonian dynamics.
Using the Hamiltonian $G$, we define 
the Poisson matrices $J^{\alpha}_{ij}(x_1^{1}, x_2^{1}, x_3^{1}, ..., x_1^{n}, x_2^{n}, x_3^{n})$ as
 \begin{eqnarray}
J^{\alpha}_{ij} \equiv \epsilon_{i j k} \frac{\partial G}{\partial x^{\alpha}_{k}}.
\label{PM1n}
\end{eqnarray}
In terms of these anti-symmetric matrices, we define the noncanonical Poisson bracket as
\begin{eqnarray}
\{A, B\}_{G}\equiv\sum_{{\alpha}=1}^{n}\{A, B\}^{\alpha}_{G}
\equiv\sum_{{\alpha}=1}^{n} J^{\alpha}_{i j} 
\frac{\partial A}{\partial x^{\alpha}_{i}}\frac{\partial B}{\partial x^{\alpha}_{j}}=\{A, B, G\},
\label{ncPB1n}
\end{eqnarray}
and then we rewrite the Nambu equation as the noncanonical Hamilton's equation of motion,
\begin{eqnarray}
\frac{d f}{dt} = \{f, H, G\}=\{f, H\}_{G}
= \sum_{{\alpha}=1}^{n} J^{\alpha}_{i j} 
\frac{\partial f}{\partial x^{\alpha}_{i}}\frac{\partial H}{\partial x^{\alpha}_{j}}.
\label{NC-eq1n}
\end{eqnarray}
Since the Nambu bracket of Eq. (\ref{NBn}) no longer satisfies the fundamental identity,
it is nontrivial whether the noncanonical Poisson bracket of Eq. (\ref{ncPB1n})
satisfies the Jacobi identity.
Let us find the conditions for the Jacobi identity to hold.
The difference between the left- and right-hand sides of Eq. (\ref{JI}) 
can be represented as
\begin{eqnarray}
{\rm lhs-rhs}=\sum_{{\alpha}=1}^{n}\sum_{{\beta}=1}^{n}\left(
\{\{A, B\}^{\alpha}_{G}, C\}^{\beta}_{G}-
\{\{A, C\}^{\alpha}_{G}, B\}^{\beta}_{G}-
\{A,\{B, C\}^{\alpha}_{G}\}^{\beta}_{G}
\right),
\label{JI1n}
\end{eqnarray}
where all the terms with $\alpha=\beta$ vanish,
because the Jacobi identity holds for each degree of freedom. 
For the terms with $\alpha\ne\beta$, 
after a straightforward calculation we obtain
\begin{eqnarray}
&&\{\{A, B\}^{\alpha}_{G}, C\}^{\beta}_{G}-
\{\{A, C\}^{\alpha}_{G}, B\}^{\beta}_{G}-
\{A,\{B, C\}^{\alpha}_{G}\}^{\beta}_{G}
~\!+\left(\alpha\leftrightarrow\beta\right)\nonumber\\
&&=\left(\partial^{\beta}_{k}J^{\alpha}_{ij}\right)J^{\beta}_{kl}
\left(
\partial^{\alpha}_{i}A~\!\partial^{\alpha}_{j}B~\!\partial^{\beta}_{l}C
-\partial^{\alpha}_{i}A~\!\partial^{\beta}_{l}B~\!\partial^{\alpha}_{j}C
+\partial^{\beta}_{l}A~\!\partial^{\alpha}_{i}B~\!\partial^{\alpha}_{j}C
\right)
+\left(\alpha\leftrightarrow\beta\right).
\label{JI2n}
\end{eqnarray}
Therefore if the Poisson matrices satisfy
\begin{eqnarray}
\frac{\partial}{\partial x^{\beta}_{k}}J^{\alpha}_{ij}=0 ~~~\left(\alpha\ne\beta\right),
\label{PM1dn}
\end{eqnarray}
then Eq. (\ref{JI2n}) becomes zero, and the Jacobi identity holds.  
\par
Consider the case that $3n$ variables are coupled in the Hamiltonian $H$,
but decoupled in the Hamiltonian $G$:
\begin{eqnarray}
G=\sum_{{\alpha}=1}^{n}G_{\alpha}(x_1^{\alpha}, x_2^{\alpha}, x_3^{\alpha}).
\label{decouple2}
\end{eqnarray}
Then the corresponding Poisson matrices of Eq. (\ref{PM1n}) are functions of the single degree of freedom,
$J^{\alpha}_{ij}=J^{\alpha}_{ij}(x_1^{\alpha}, x_2^{\alpha}, x_3^{\alpha})$,
and satisfy the condition of Eq. (\ref{PM1dn}), and the Jacobi identity holds.
In this case the consistent time evolution is broken in the original Nambu dynamics,
but restored in the corresponding noncanonical Hamiltonian dynamics.
On the other hand, if we define the Poisson matrices by means of the Hamiltonian $H$,
\begin{eqnarray}
\tilde{J}^{\alpha}_{ij} \equiv -\epsilon_{i j k} \frac{\partial H}{\partial x^{\alpha}_{k}},
\label{PM2n}
\end{eqnarray}
and rewrite the Nambu equation as
\begin{eqnarray}
\frac{d f}{dt} = \{f, H, G\}=\{f, G\}_{H}
= \sum_{{\alpha}=1}^{n} \tilde{J}^{\alpha}_{i j} 
\frac{\partial f}{\partial x^{\alpha}_{i}}\frac{\partial G}{\partial x^{\alpha}_{j}},
\label{NC-eq2n}
\end{eqnarray}
then the Jacobi identity does not hold, and  
the consistent time evolution cannot be restored.
This is because $3n$ variables are not decoupled in the Hamiltonian $H$,
and therefore $H$ cannot be written in the decomposed form,
$H=\sum_{{\alpha}=1}^{n}H_{\alpha}(x_1^{\alpha}, x_2^{\alpha}, x_3^{\alpha})$.
It should be noted that the Liouville theorem holds in the dynamics of both 
Eqs. (\ref{NC-eq1n}) and (\ref{NC-eq2n}).

\subsection{Example: Semiclassical coupled oscillators}
\label{Example}
As an example of many degrees of freedom systems,
consider a 1D system of two quantum oscillators whose Hamiltonian is given by
\begin{eqnarray}
\hat{H}=
\frac{1}{2m_{1}}\hat{p}_{1}^{2}+\frac{1}{2m_{2}}\hat{p}_{2}^{2}
+\frac{m_{1}\omega_{1}^{2}}{2}\hat{q}_{1}^{2}
+\frac{m_{2}\omega_{2}^{2}}{2}\hat{q}_{2}^{2}
+\lambda~\!\hat{q}_{1}\hat{q}_{2}^{2}.
\label{sHH}
\end{eqnarray}
This is a simplified version of the quantum H\'enon--Heiles model \cite{HellerStechelDavis}.
The semiclassical equations of motion for the quantum expectation values 
$(\langle\hat{q}_{1}\rangle, \langle\hat{p}_{1}\rangle, \langle\hat{q}^{2}_{1}\rangle,
\langle\hat{q}_{2}\rangle, \langle\hat{p}_{2}\rangle, \langle\hat{q}^{2}_{2}\rangle)$
are given by approximating the higher-order expectation values by means of the lower ones \cite{PrezhdoPereverzev}:
\begin{eqnarray}
&&
\frac{d}{dt}\langle\hat{q}_{1}\rangle=\frac{1}{m_{1}}\langle\hat{p}_{1}\rangle,~~~~~~~~~~~~~~~~~~~~~~
\frac{d}{dt}\langle\hat{q}_{2}\rangle=\frac{1}{m_{2}}\langle\hat{p}_{2}\rangle,
\nonumber\\
&&
\frac{d}{dt}\langle\hat{p}_{1}\rangle=
-m_{1}\omega_{1}^{2}\langle\hat{q}_{1}\rangle-\lambda \langle\hat{q}_{2}^{2}\rangle,~~~~~~~\!
\frac{d}{dt}\langle\hat{p}_{2}\rangle\simeq
-m_{2}\omega_{2}^{2}\langle\hat{q}_{2}\rangle-2\lambda\langle\hat{q}_{1}\rangle\langle\hat{q}_{2}\rangle,
\nonumber\\
&&
\frac{d}{dt}\langle\hat{q}_{1}^{2}\rangle\simeq
\frac{2}{m_{1}}\langle\hat{q}_{1}\rangle\langle\hat{p}_{1}\rangle,~~~~~~~~~~~~~~~~~
\frac{d}{dt}\langle\hat{q}_{2}^{2}\rangle\simeq
\frac{2}{m_{2}}\langle\hat{q}_{2}\rangle\langle\hat{p}_{2}\rangle.
\label{SC-eqs}
\end{eqnarray} 
This semiclassical dynamics can be formulated as the Nambu dynamics 
using the hidden Nambu formalism \cite{HorikoshiKawamura,Horikoshi}.
We choose $n=2$, $N=3$ Nambu variables as follows:
\begin{eqnarray}
\left(
    \begin{array}{c}
      x_1^{1}\\
      x_2^{1}\\
      x_3^{1}
    \end{array}
  \right)
=
\left(
    \begin{array}{c}
      \langle\hat{q}_{1}\rangle \\
      \langle\hat{p}_{1}\rangle \\
      \langle\hat{q}_{1}^{2}\rangle
    \end{array}
  \right),~~~
\left(
    \begin{array}{c}
      x_1^{2}\\
      x_2^{2}\\
      x_3^{2}
    \end{array}
  \right)
=
\left(
    \begin{array}{c}
      \langle\hat{q}_{2}\rangle \\
      \langle\hat{p}_{2}\rangle \\
      \langle\hat{q}_{2}^{2}\rangle
    \end{array}
  \right),
\label{Exv}
\end{eqnarray}
and define the Nambu Hamiltonians $H$ and $G$ as
\begin{eqnarray}
H&=&\frac{1}{2m_{1}}\left(x_2^{1}\right)^{2}+\frac{1}{2m_{2}}\left(x_2^{2}\right)^{2}
+\frac{m_{1}\omega_{1}^{2}}{2}x_3^{1}+\frac{m_{2}\omega_{2}^{2}}{2}x_3^{2}
+\lambda~\!x_1^{1}x_3^{2},
\label{ExH}\\
G&=&\sum_{{\alpha}=1}^{2}\left(x_3^{\alpha}-\left(x_1^{\alpha}\right)^2\right).
\label{ExG}
\end{eqnarray}
Then it can be shown that the Nambu equation (\ref{N-eqn}) reproduces the 
semiclassical equations (\ref{SC-eqs}). 
This is a semiclassical dynamics with constraints that the quantum fluctuation of each mode,
$\langle\hat{q}_{\alpha}^{2}\rangle-\langle\hat{q}_{\alpha}\rangle^{2}$, is constant in time.
Therefore the Hamiltonian $G$ can be written in the decomposed form of Eq. (\ref{ExG}).
Since the Hamiltonian $H$ has an interaction term between two degrees of freedom,
the fundamental identity for these $H$ and $G$ does not hold \cite{Horikoshi}.
For example, if we choose $(A, B, C)=(x_{1}^{2}, x_{2}^{2}, x_{2}^{1})$ and $(D, E)=(H, G)$,
then the left-hand side of Eq. (\ref{FI}) is zero, whereas the right-hand side is $-\lambda$.
This implies that the consistent time evolution is broken in this Nambu dynamics.
\par
Let us see if the Jacobi identity holds in two corresponding types of noncanonical Hamiltonian dynamics. 
First, if we define the Jacobi matrices using the Hamiltonian $G$ as in Eq. (\ref{PM1n}),
they can be written as
\begin{eqnarray}
J^{1} = \left(
    \begin{array}{ccc}
      0 & 1 & 0 \\
      -1 & 0 & -2x^{1}_{1} \\
      0 & 2x^{1}_{1} & 0
    \end{array}
  \right),~~~
J^{2} = \left(
    \begin{array}{ccc}
      0 & 1 & 0 \\
      -1 & 0 & -2x^{2}_{1} \\
      0 & 2x^{2}_{1} & 0
    \end{array}
  \right).
\label{ExPM1}
\end{eqnarray}
These satisfy Eq. (\ref{PM1dn}), and therefore the Jacobi identity holds.
For example, if we choose $(A, B)=(x_{2}^{1}, x_{2}^{2})$ and $C=H$,
then both sides of Eq. (\ref{JI}) are zero.
On the other hand, if we define the Jacobi matrices in another way using the Hamiltonian $H$ as in Eq. (\ref{PM2n}),
they read
\begin{eqnarray}
\tilde{J}^{1} = \left(
    \begin{array}{ccc}
      0 & -\frac{m_{1}\omega^{2}_{1}}{2} & \frac{1}{m_{1}}x^{1}_{2} \\
      \frac{m_{1}\omega^{2}_{1}}{2} & 0 & -\lambda x^{2}_{3} \\
      -\frac{1}{m_{1}}x^{1}_{2} & \lambda x^{2}_{3} & 0
    \end{array}
  \right),~~
\tilde{J}^{2} = \left(
    \begin{array}{ccc}
      0 & -\frac{m_{2}\omega^{2}_{2}}{2}-\lambda x^{1}_{1} & \frac{1}{m_{2}}x^{2}_{2} \\
      \frac{m_{2}\omega^{2}_{2}}{2}+\lambda x^{1}_{1} & 0 & 0 \\
      -\frac{1}{m_{2}}x^{2}_{2} & 0 & 0
    \end{array}
  \right).
\label{ExPM2}
\end{eqnarray}
These do not satisfy Eq. (\ref{PM1dn}), and therefore the Jacobi identity is violated.
If we choose $(A, B)=(x_{2}^{1}, x_{2}^{2})$ and $C=H$ again,
then the left-hand side of Eq. (\ref{JI}) is zero, whereas the right-hand side is 
$-\lambda m_{1}\omega^{2}_{1}x^{2}_{1}$.
The consistent time evolution is restored in the noncanonical Hamiltonian dynamics 
with the Poisson matrices of Eq. (\ref{ExPM1}), but remains broken in the dynamics 
with Eq. (\ref{ExPM2}).

\section{Conclusions}
\label{Conclusions}
It is well known that the Nambu bracket does not satisfy the fundamental identity in many degrees of freedom systems.
In the present paper, we have shown that 
the noncanonical Poisson bracket defined by the Nambu bracket could satisfy the Jacobi identity,
and derived the condition for it, Eq. (\ref{PM1dn}). 
We have given an example of a two degrees of freedom system to show
the breaking and restoration of the consistent time evolution in the Nambu dynamics
and the corresponding noncanonical Hamiltonian dynamics.
\par
The violation of the Jacobi identity 
is an important subject in the generalized Hamiltonian dynamics \cite{SatoYoshida,Sato}.
As for the fundamental identity, its violation in many degrees of freedom systems
implies a difficulty with formulating the statistical mechanics of Nambu variables.
Therefore it would be interesting to see if we could construct 
effective statistical mechanics of Nambu variables
by means of the noncanonical Hamiltonian representation or its analogs.

\section*{Acknowledgments}
This work was partly supported by Osaka City University Advanced Mathematical Institute
 (MEXT Joint Usage/Research Center on Mathematics and Theoretical Physics JPMXP0619217849).


\end{document}